\documentclass[11pt, a4paper, submission, Proceedings]{SciPost}
\pdfoutput=1

\hypersetup{
    colorlinks,
    linkcolor={red!50!black},
    citecolor={blue!50!black},
    urlcolor={blue!80!black}
}

\usepackage[bitstream-charter]{mathdesign}
\DeclareSymbolFont{usualmathcal}{OMS}{cmsy}{m}{n}
\DeclareSymbolFontAlphabet{\mathcal}{usualmathcal}

\begin{document}

\begin{center}{\Large \textbf{
CORSIKA 8 -- the next-generation air shower simulation framework\\
}}\end{center}

\begin{center}
Tim Huege\textsuperscript{1,2,$\star$} for the CORSIKA 8 Collaboration\footnote{\url{https://gitlab.iap.kit.edu/AirShowerPhysics/corsika/-/wikis/CORSIKA-Talks/ICRC2021-author-list}}
\end{center}

\begin{center}
{\bf 1} Karlsruhe Institute of Technology, Institute for Astroparticle Physics (IAP),\\ P.O.\ Box 3640, 76021 Karlsruhe, Germany
\\
{\bf 2} Vrije Universiteit Brussel, Astrophysical Institute,\\ Pleinlaan 2, 1050 Brussels, Belgium
\\
* tim.huege@kit.edu
\end{center}

\begin{center}
\today
\end{center}


\definecolor{palegray}{gray}{0.95}
\begin{center}
\colorbox{palegray}{
  \begin{tabular}{rr}
  \begin{minipage}{0.1\textwidth}
    \includegraphics[width=30mm]{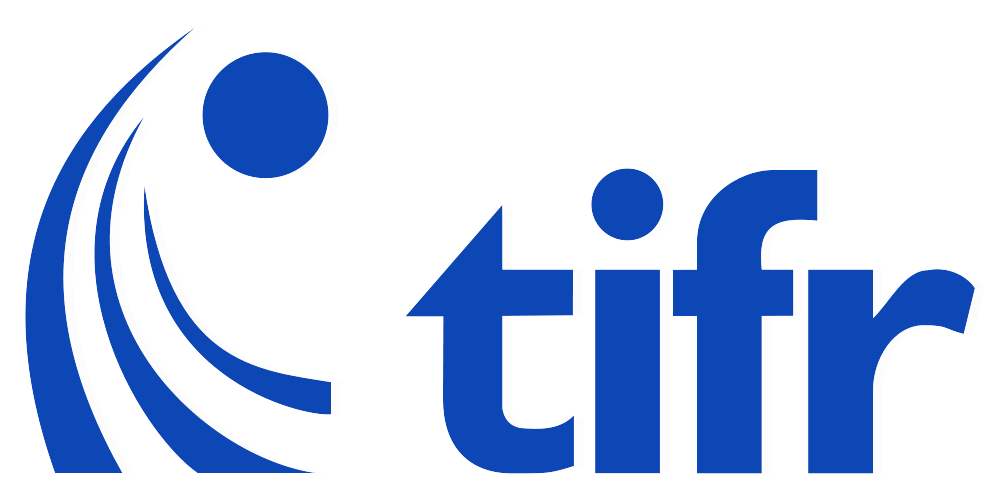}
  \end{minipage}
  &
  \begin{minipage}{0.85\textwidth}
    \begin{center}
    {\it 21st International Symposium on Very High Energy Cosmic Ray Interactions (ISVHE- CRI 2022)}\\
    {\it Online, 23-27 May 2022} \\
    \doi{10.21468/SciPostPhysProc.?}\\
    \end{center}
  \end{minipage}
\end{tabular}
}
\end{center}

\section*{Abstract}
{\bf
For more than 20 years, the community has heavily relied on CORSIKA for the simulation of extensive air showers, their Cherenkov light emission and their radio signals. While tremendously successful, the Fortran-based monolithic design of CORSIKA up to version 7 limits adaptation to new experimental needs, for example, in complex scenarios where showers transition from air into dense media, and to new computing paradigms such as the use of multi-core and GPU parallelization. With CORSIKA 8, we have reimplemented the core functionality of CORSIKA in a modern, modular, C++-based simulation framework, and successfully validated it against CORSIKA 7. Here, we discuss the philosophy of CORSIKA 8, showcase some example applications, and present the current state of implementation as well as the plans for the future.
}


\section{Introduction}
\label{sec:intro}

For more than 20 years, CORSIKA \cite{HeckKnappCapdevielle1998} has been \emph{the} workhorse Monte Carlo simulation code used by the community to simulate extensive air showers as well as their Cherenkov light and radio emission. When CORSIKA was developed, originally for the KASCADE experiment \cite{KASCADE:2003swk}, it finally provided the community with a common reference that everybody could compare and relate to for their simulation needs, and was thus received very well by the community. Meticulous documentation and continuous maintenance as well as fast updates in numerous releases\footnote{At the time of writing, the current version is 7.7420.} to include the newest interaction models and added functionalities were further cornerstones for CORSIKA's success. The widespread use of CORSIKA is clearly illustrated by the fact that the corresponding report \cite{HeckKnappCapdevielle1998} has hitherto gathered more than 1,000 citations.

With the advancement in computing infrastructures, in particular the trend to increase the number of CPU cores rather than the performance of single cores and the widespread availability of GPU clusters, however, the need for a more modern code design became increasingly clear. Also, the needs of complex experimental setups such as those measuring cross-media showers starting in air and then transitioning into ice clash with the limitations of the monolithic and hand-optimized code design of CORSIKA 7 based on Fortran in various dialects that the current generation of researchers is no longer familiar with.

In 2018, we have thus begun a complete re-implementation of the core functionality of CORSIKA in a modern, C++-based simulation framework \cite{Engel:2018akg}. In this article, we describe the philosphy and give a concise overview of the current state of this \emph{CORSIKA 8} project. We note that the results reported here summarize the state reached at the time of the ICRC2021, while in the meantime work has naturally been progressing and further updates are under preparation.


\section{Philosophy and code design}

The goal of the CORSIKA 8 project is to provide a \emph{modular} and thus flexible re-implementation of the CORSIKA 7 (core) functionality in a modern C++-based code framework. The modularity will facilitate the development of new capabilities without users having to navigate the complete code, as the interfaces are well-defined and new algorithms can be plugged in in a defined way. Karlsruhe Institute of Technology is committed to provide long-term development and maintenance for the core functionalities, yet unlike CORSIKA 7 the project is intended as a ``community effort'' channeling the expertise of authors from different experimental projects.

\begin{figure}[htb]
\centering
\includegraphics[width=1.0\textwidth]{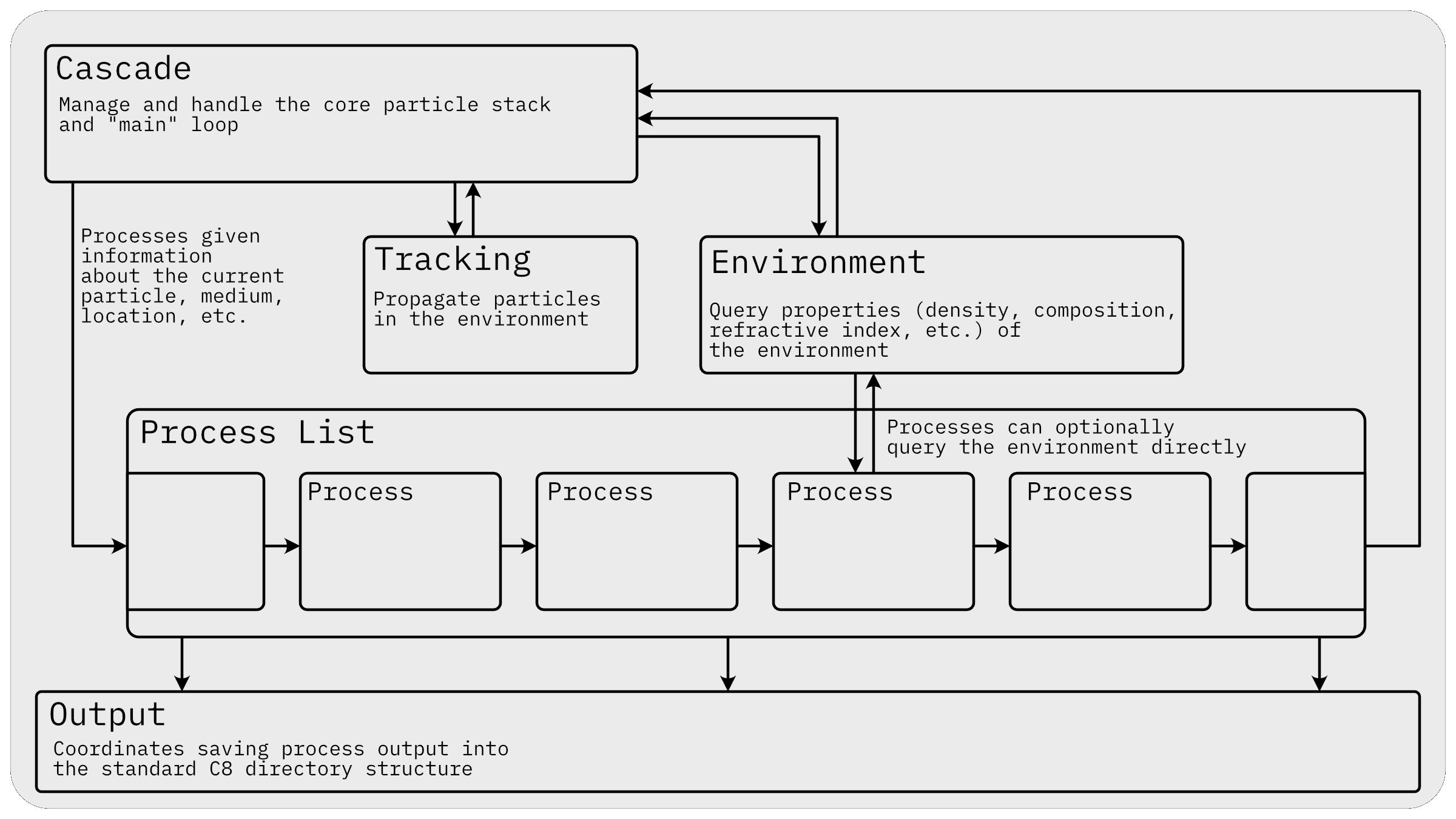}
\caption{General structure of the CORSIKA 8 code, see text for further description.}
\label{fig:flowchart}
\end{figure}

The general structure of the code is visualized in figure \ref{fig:flowchart}. At the heart of the simulation is the \emph{Cascade} which loops over the particles on the stack, lets them be \emph{tracked} in the \emph{Environment} and passes them to the \emph{Process List}. This \emph{Process List} can be flexibly arranged by the user and will host a mixture of discrete processes (e.g., collisions, decays, interactions) and continuous processes (e.g., energy loss, multiple scattering, Cherenkov-light and radio-emission calculations). A further kind of process is the crossing of a \emph{boundary} in the environment. A common \emph{Output Handler} will take care of writing information out as needed.

For performance reasons, we employ compile-time optimization where possible and avoid run-time virtualization. We rely heavily on templating and try to minimize dependencies on external libraries. A continuous integration workflow is performing unit tests and physics validation during collaborative development. The code is openly available on our GitLab repository\footnote{\url{https://gitlab.iap.kit.edu/AirShowerPhysics/corsika}} with instructions on the associated Wiki on how to contribute to the project.

A strong asset of CORSIKA 8 when compared to CORSIKA 7 is the flexibility in defining the \emph{Enviroment}, which consists of an arrangement of geometrical objects (spheres, cubes, ...) that can each have individual media properties (e.g., air, ice, ..., homogeneous or with a density gradient). Geometric calculations are handled transparently based on functionality originally developed for the Offline framework of the Pierre Auger Collaboration \cite{Argiro:2007qg,Dembinski:2020wrp}. Physical units are handled by a header-only class that associates a dimension to every variable representing a physical quantity and will report illegal computations already at compile time \cite{Dembinski:2020wrp}. Efficient use of parallelization on CPUs and/or GPUs is foreseen in the later development. Further details on the general structure and design of the code can also be found in ref.\ \cite{Reininghaus:2019jxg}.

CORSIKA 8 is already able to simulate complete extensive air showers, including hadronic and electromagnetic interactions. Simulation steering is currently still somewhat rudimentary and the final implementation has not yet been defined. Output is currently written in a combination of human-readable YAML files and Apache Parquet files\footnote{\url{https://parquet.apache.org/}} for binary data, with an associated Python library provided for easy read-in by users. However, the output format might still be changed based on experience made with the current choice. In general, the code is ready for expert-level users and interested developers, yet needs some more work to become ready for end users.

In the following, we will describe the available functionality and illustrate some CORSIKA 8 results in comparison with reference simulations with CORSIKA 7 and other codes.


\section{Hadronic interactions}

A wide range of current hadronic interaction models is already available within CORSIKA 8. At high energies, in addition to QGSJET-II.04 \cite{PhysRevD.83.014018}, Sybill 2.3d \cite{Engel:2019dsg} and EPOS-LHC \cite{Pierog:2013ria}, also Pythia 8 \cite{Sjostrand:2021dal} is included and can thus be used for air-shower simulations for the first time. At low energies, currently UrQMD \cite{Bleicher:1999xi} is available with a plan to also include FLUKA \cite{Ferrari:2005zk}. Decays can be handled by Sybill 2.3d and Pythia 8.

\begin{figure}[htb]
\centering
\includegraphics[width=1.0\textwidth]{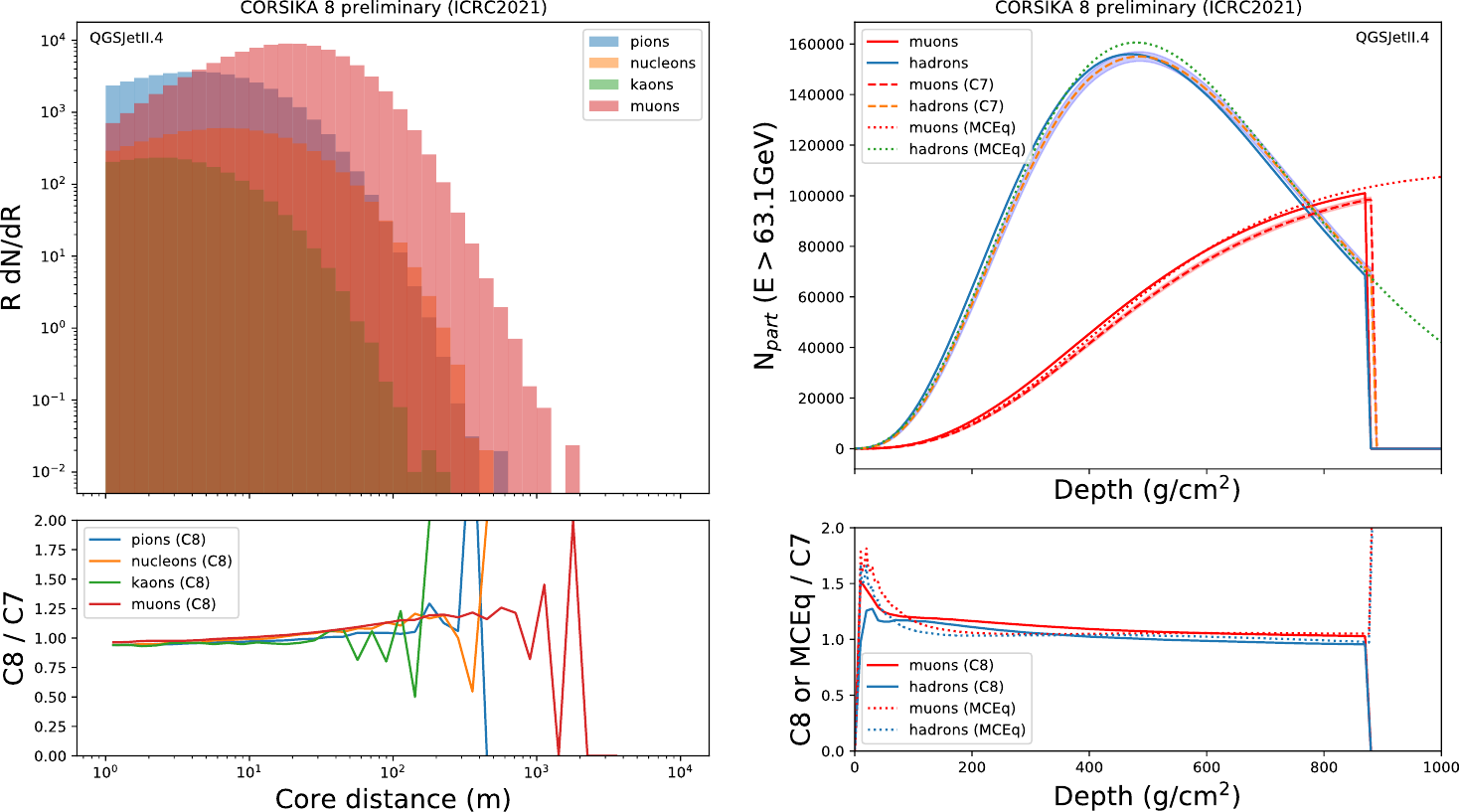}
\caption{Comparison of the hadronic particle cascade in air showers simulated with CORSIKA 8 and CORSIKA 7 as well as MCEq \cite{Fedynitch:2015zma}. Particles with energies above 63.1~GeV  are shown (the lower limit of 10 GeV center-of-mass energy that Sibyll 2.3d can handle). The left figure illustrates the lateral distributions of particles while the right figure shows the longitudinal shower evolution profiles. From \cite{CORSIKA8:2021gjr}.}
\label{fig:hadroncascade}
\end{figure}

We have performed in-depth comparisons of hadronic cascades with simulations with CORISKA 7 and MCEq, as illustrated in figure \ref{fig:hadroncascade}. Both the lateral particle distributions (left) and the air-shower evolution of the particles in the hadronic cascade are in good agreement with CORSIKA 7 and MCEq, on the level of typically 10\%. These studies also helped to discover some (minor) problems in earlier versions of CORSIKA 7 which were fixed in follow-up releases. Further refinements and validation work in CORSIKA 8 are ongoing and will be reported at a later point. A feature worth mentioning here is also the availability of tracking ``mother particles'' in CORSIKA 8 allowing to do in-depth air-shower geneaology \cite{Reininghaus:2021zge} and thus study air-shower physics in great detail, in particular to better understand the ``muon puzzle''.


\section{Electromagnetic interactions}

Electromagnetic interactions in CORSIKA 8 are being handled by the PROPOSAL \cite{Dunsch:2018nsc,Alameddine:2020zyd} code, while CORSIKA 7 uses a modified version of EGS4 \cite{Nelson:1985ec}. All relevant interactions treated in EGS are also handled by PROPOSAL, with a few limitations that still need to be addressed. The LPM effect in CORSIKA 8 is currently only applicable to homogeneous media, but is in principle available in PROPOSAL and will shortly be taken into account also for inhomogeneous media. Photohadronic interactions at energies well below 100~GeV still need to be included. Particles down to an energy of 2~MeV can be fully treated, for radio-emission calculations, however, this threshold will need to be lowered to 500~keV. The photo effect is not yet taken into account but would be relevant mostly at yet lower energies.

\begin{figure}[htb]
\centering
\includegraphics[width=1.0\textwidth]{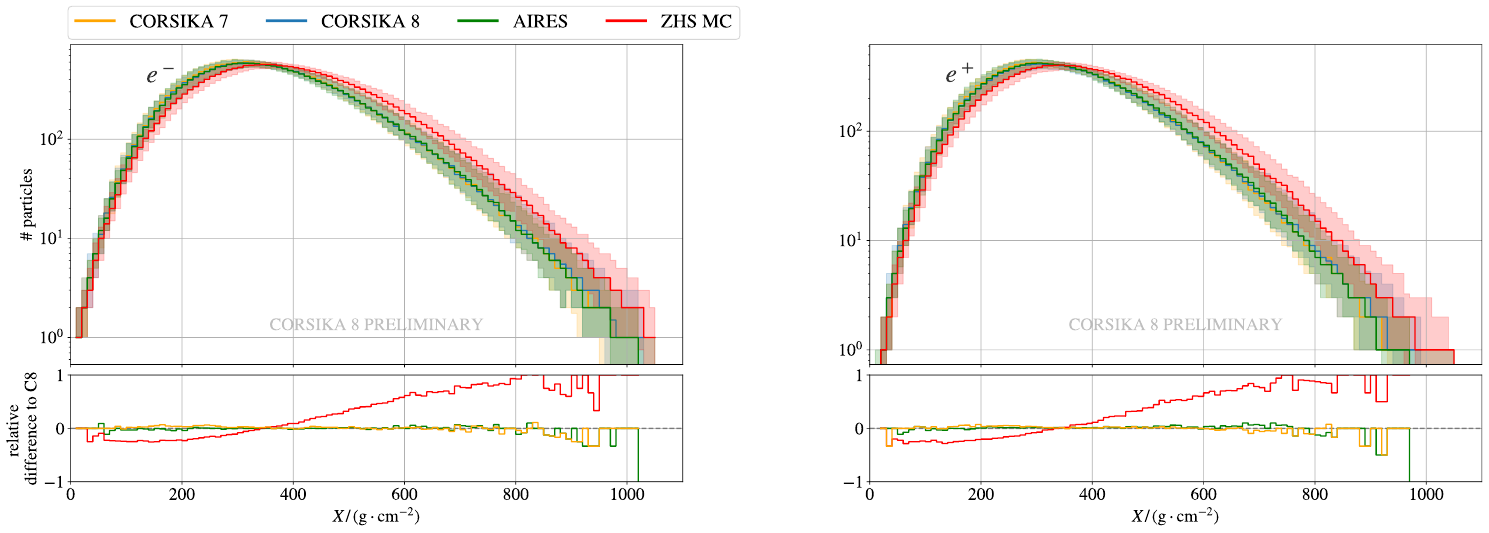}
\caption{Comparison of the longitudinal evolution of the electron and positron content for particles with an energy above 4~MeV in CORSIKA 8 with results from CORSIKA 7, AIRES \cite{Sciutto1999} and the ZHS Monte Carlo code \cite{Zas:1991jv}. The lines show the mean over 200 showers started by a 1~TeV electron, the shaded regions illustrate the interquartile (middle 50\%) range. From \cite{CORSIKA8:2021ilo}.}
\label{fig:emcascade}
\end{figure}

\begin{figure}[htb]
\centering
\includegraphics[width=0.8\textwidth]{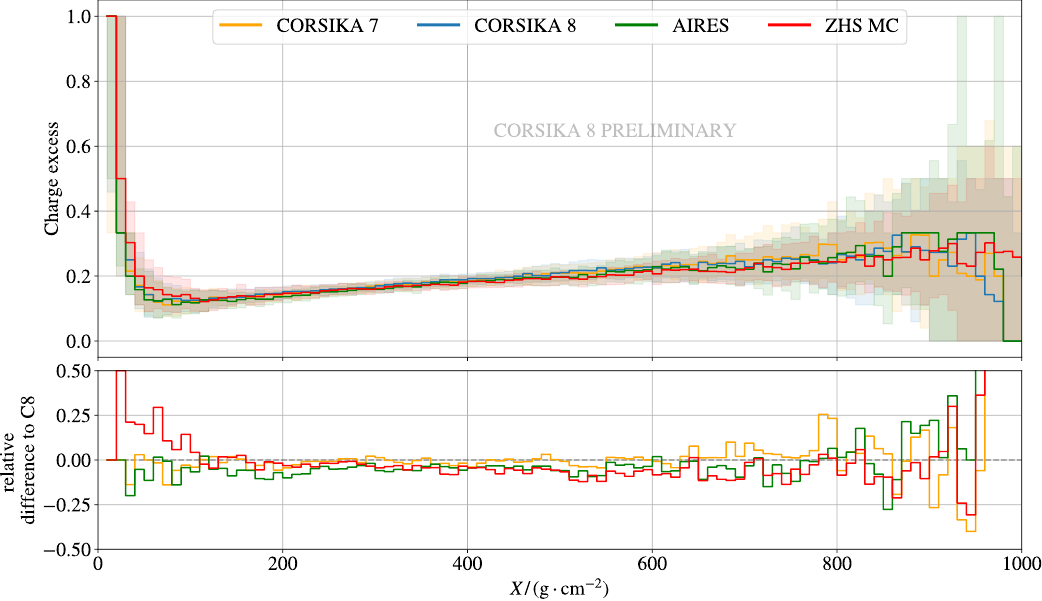}
\caption{Comparison of the longitudinal evolution of the charge excess for particles above an energy of 4~MeV in CORSIKA 8 with results from CORSIKA 7, AIRES and the ZHS Monte Carlo code. Same showers and definitions as used in figure \ref{fig:emcascade}. From \cite{CORSIKA8:2021ilo}.}
\label{fig:chargeexcess}
\end{figure}

Again, in-depth comparisons have been performed with CORSIKA 7 and other codes. As illustrated in figure \ref{fig:emcascade}, there is good agreement, at the level of 5\%, in the longitudinal evolution of electrons and positrons between CORSIKA 8, CORSIKA 7 and AIRES. Also the lateral distributions have been checked and shown to be in similar agreement. Figure \ref{fig:chargeexcess} illustrates the comparison of the charge excess in the air shower as a function of the longitudinal evolution of the electromagnetic cascade. Again, there is agreement at the level of better than 10\% with CORSIKA 7 and AIRES. The charge excess in air showers is relevant, in particular, for the simulation of radio emission.


\section{Beyond air showers}

\begin{figure}[htb]
\centering
\includegraphics[width=0.7\textwidth]{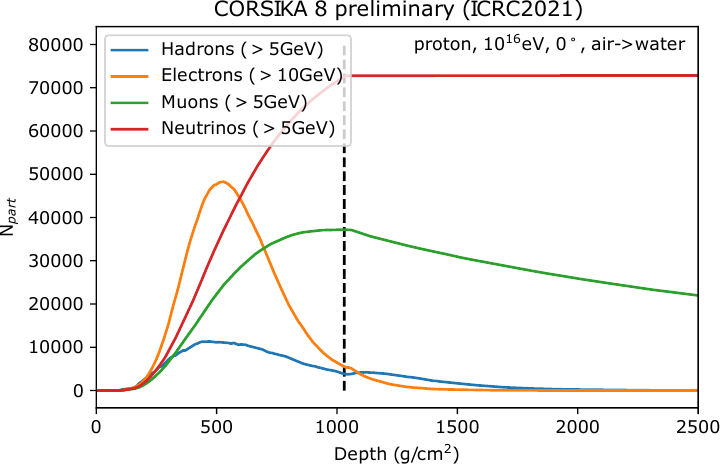}
\caption{Simulation of an air shower transitioning into water as simulated with CORSIKA 8. From \cite{CORSIKA8:2021gjr}.}
\label{fig:crossmedia}
\end{figure}

There is a strong need in the community to simulate particle showers starting in air and then propagating into dense media, in particular ice, for example, to determine the potential background for in-ice radio detectors for high-energy neutrinos. Existing codes, in particular CORSIKA 7, cannot handle this scenario due to limitations in defining environmental properties. In CORSIKA 8, on the other hand, the flexible environment definition allows us to simulate such cross-media showers already now, as illustrated in figure \ref{fig:crossmedia} for an air shower passing over into water. Other media such as the Martian atmosphere or solids can also easily be employed, as long as they can be handled adequately by the underlying interaction models. Further details can be found in ref.\ \cite{CORSIKA8:2021gjr}.

\section{Electromagnetic emission from particle showers}

Not only the particle cascades themselves are of interest in CORSIKA simulations, but also electromagnetic emission by the cascade, in particular radio emission and Cherenkov light.

\subsection{Radio-emission calculation}

The radio-detection community perhaps has the strongest need to transition to a more modern simulation framework than CORSIKA 7, for example, because they need to simulate cross-media showers but also upgoing and Earth-skimming geometries as well as showers the radio signal of which is reflected off the Earth's surface. CORSIKA 8 will give us the opportunity to handle all of these different cases consistently within one common code, greatly reducing complexity and maintenance in comparison with using multiple codes.

\begin{figure}[h!tb]
\centering
\includegraphics[width=1.0\textwidth]{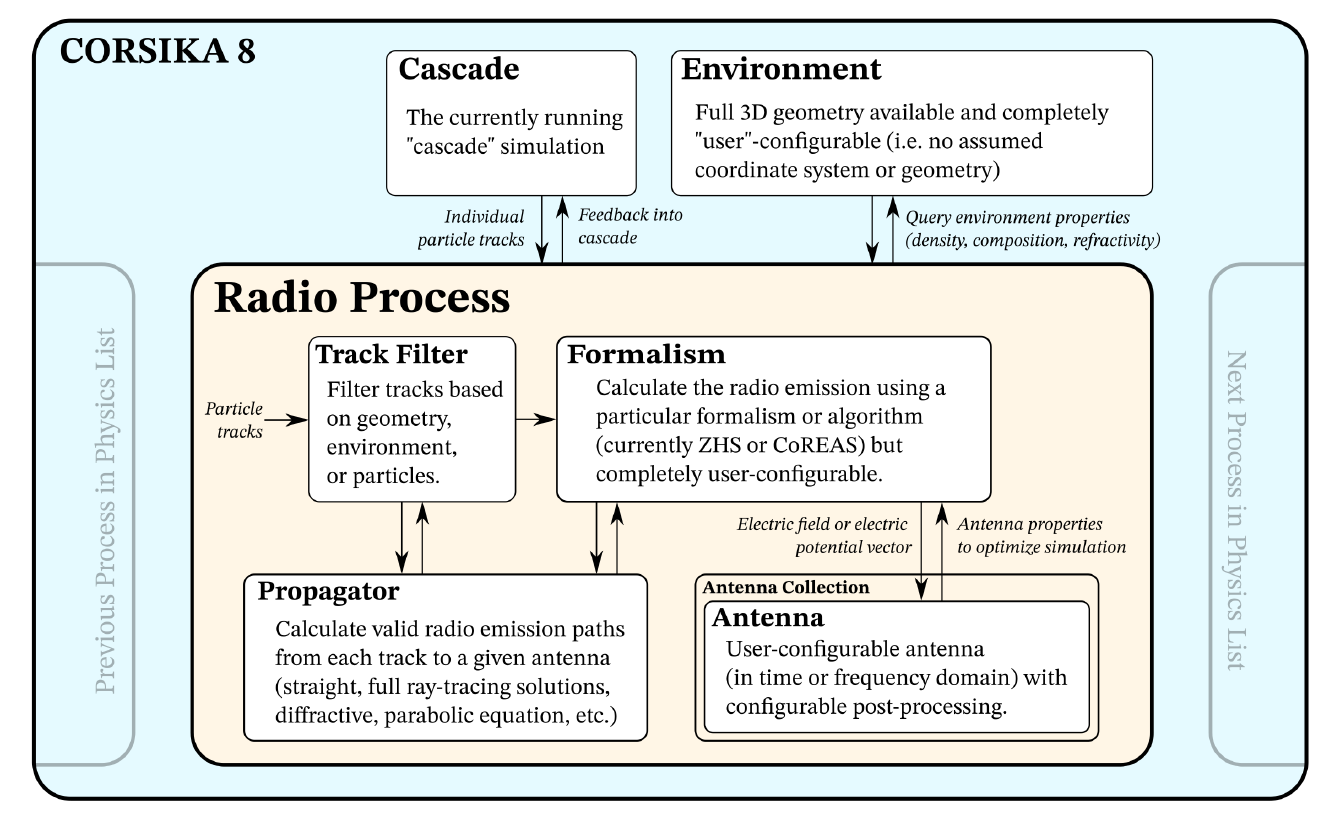}
\caption{Structure of the \emph{Radio Process} in CORSIKA 8. Please see text for further explanations. From \cite{Karastathis:2021akf}.}
\label{fig:radiomodule}
\end{figure}

Radio-emission calculations have been implemented in CORSIKA 8 taking full advantage of the modularity that the code offers. They are performed in just another process that can be added to the \emph{Process List}. The structure of this \emph{Radio Process} is illustrated in figure \ref{fig:radiomodule}. Internally, it is split in four sub-components: A \emph{Track Filter} allows to discard particles according to user-specified criteria (e.g., particle energy or shower depth). The \emph{Antenna Collection} hosts user-configurable \emph{Antennas} which can record time- or frequency-domain signals and can potentially already apply an antenna model on-the-fly. Signal propagation from the particle positions to the antennas is handled in the \emph{Propagator} which can be user-provided to do anything from simple straight-line propagation up to raytracing in complex media, as will become necessary for the proper simulation of radio signals in polar ice and firn. Finally, alternative \emph{Formalisms} for the calculation of the electromagnetic radiation from the particle tracks can be implemented and activated -- even in parallel. For the moment, 1:1 ports of the ``endpoints formalism'' in CoREAS \cite{Huege:2013vt} (in CORSIKA 7) and the ZHS formalism as implemented in ZHAireS \cite{AlvarezMunizCarvalhoZas2012} are available in CORSIKA 8.

\begin{figure}[h!tb]
\centering
\includegraphics[width=1.0\textwidth,trim={3 3 3 3},clip]{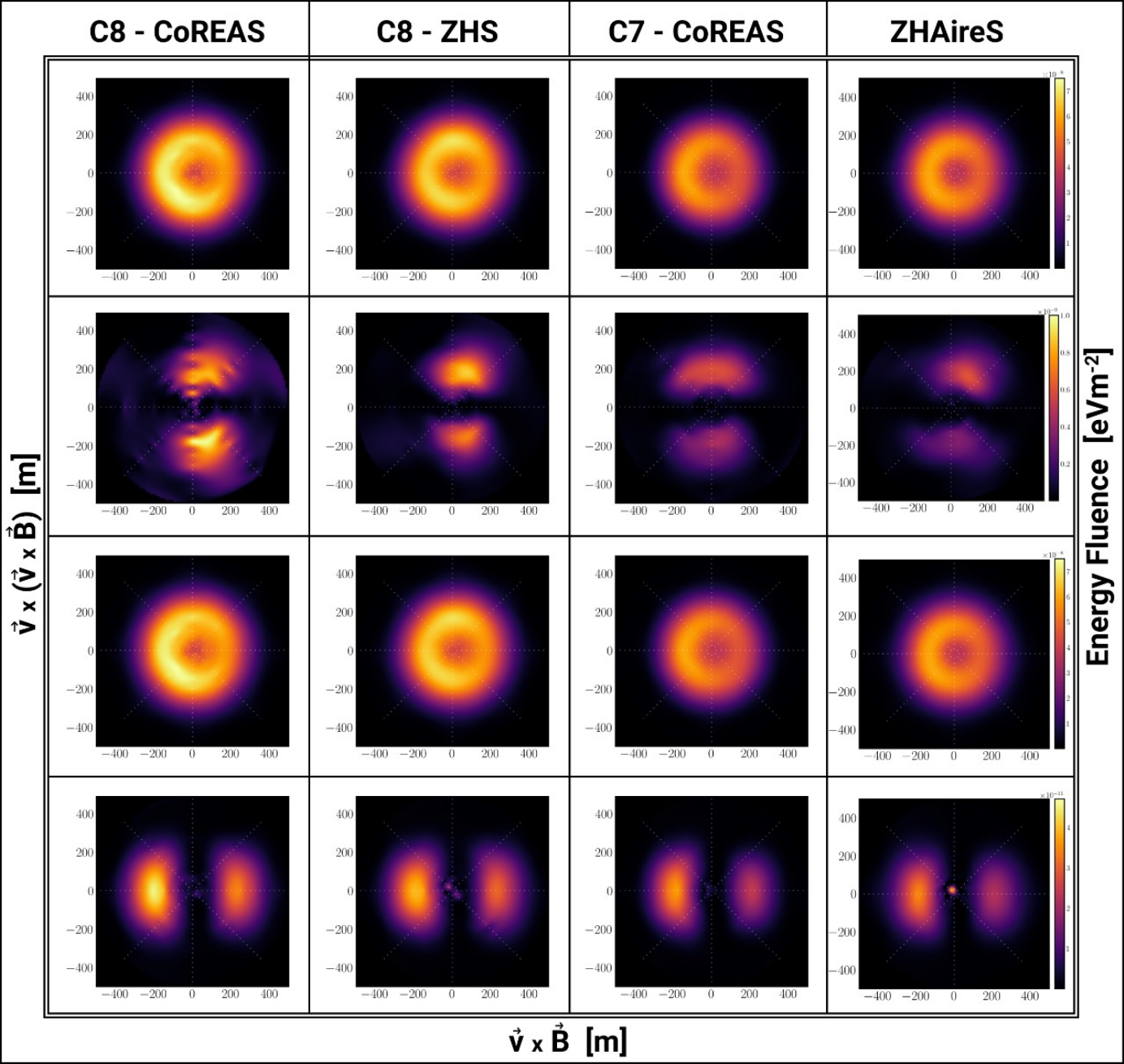}
\caption{Maps of the energy fluence in the 30--80~MHz band as simulated with different formalisms in different codes for a vertical, electron-induced air shower with 10~TeV of energy. The rows show, from top to bottom, the total fluence and the fluences in the $\vec{v} \times (\vec{v} \times \vec{B})$ (north-south), $\vec{v} \times \vec{B}$ (east-west) and $\vec{v}$ (vertical) polarizations. The x-axes show core distances along the $\vec{v} \times \vec{B}$ (east-west) direction and the y-axes denote core distances along the $\vec{v} \times (\vec{v} \times \vec{B})$ (north-south) direction. The color scales are identical within a given row. From \cite{Karastathis:2021akf}.}
\label{fig:radiomaps}
\end{figure}

We have compared the radio-emission simulations using both the CoREAS and ZHS formalisms in CORSIKA 8 (for the exact same shower) with simulations using CoREAS in CORSIKA 7 and ZHS in ZHAireS (showers selected to have similar longitudinal evolution). Figure \ref{fig:radiomaps} shows energy fluence maps in the 30--80~MHz band for a vertical electron-induced shower with 10~TeV of energy. The atmosphere corresponds to ``US standard'' but with a uniform refractive index of $n = 1.000327$. The geomagnetic field is chosen horizontal with a value of 50~$\mu$T. Please note that particles with energies below 5~MeV were cut in the simulation. Overall, there is good agreement between the fluence maps of the different formalisms and codes in all the polarizations. On closer look, differences are visible such as the more pronounced asymmetry in the third row, which shows the east-west ($\vec{v} \times \vec{B}$) polarization of the signal. Also, in the other polarizations, small differences can be observed. More quantitative comparisons of pulse shapes and frequency spectra are available in ref.\ \cite{Karastathis:2021akf}.

Radio signals from air showers are very sensitive to details of the particle distributions (energy, spatial, charge imbalance, ...) due to their coherent nature. They can thus be used as a powerful diagnostic, and we are currently working on studying the visible differences in more detail to understand their exact origin.

\subsection{Cherenkov-light calculation}

Another important application for CORSIKA is the calculation of Cherenkov light, needed in the context of both imaging and non-imaging Cherenkov telescopes. Currently, two approaches have been followed to provide simulations of Cherenkov light in CORSIKA 8. One is based on a port of the existing CORSIKA 7 functionality to CORSIKA 8 \cite{Carrere:2021rty}. The other explores the potential performance gain achievable with GPU parallelization by exporting particles from CORSIKA 8 into a python-based Cherenkov light calculation leveraging the power of GPUs for signal propagation \cite{Reininghaus:2021qoa}. In the future, we envisage inclusion of Cherenkov light calculation capabilities in CORSIKA 8 along similar lines and structures as in the \emph{Radio Process}; in fact, the same structure (with different implementations) should lend itself well also to the calculation of Cherenkov light.


\section{Conclusions and Outlook}

Started in 2018, the CORSIKA 8 project has by now made great strides to becoming a new, flexible and modern Monte Carlo simulation framework for particle showers in air or dense media, including their emissions in the optical and radio regime. While some work is still needed before a release that can be used by end users, the vast majority of the core functionality is already present. Hadronic and electromagnetic cascades have been shown to agree with CORSIKA 7 and other codes, and also the radio-emission and Cherenkov-light simulation functionality has been implemented and validated. The state presented here corresponds to the one shown at the ICRC2021. In the meantime, more work has been done and further results will be published in due course. If all goes well, a beta release by the ICRC2023 seems feasible. Expert users are already invited to contribute to the project through our GitLab repository.


\section*{Acknowledgements}

The authors acknowledge support by the High Performance and Cloud Computing Group at the Zentrum für Datenverarbeitung of the University of Tübingen, the state of Baden-Württemberg through bwHPC
and the German Research Foundation (DFG) through grant no INST 37/935-1 FUGG.







\nolinenumbers

\end{document}